\documentclass[twocolumn,showpacs,aps,pra]{revtex4}

\usepackage{epsfig}

\begin{document}

\def\br{{\bf r}}
\def\bk{{\bf k}}
\def\beq{\begin{equation}}
\def\eeq{\end{equation}}
\def\bea{\begin{eqnarray}}
\def\eea{\end{eqnarray}}
\def\bp{{\bf p}}
\def\bv{{\bf v}}
\def\be{\eta}
\newcommand{\eq}[1]{\mbox{(\ref{#1})}}

\title{Optimal conditions for observing Josephson oscillations in a double-well Bose-gas condensate}

\author{J. E. Williams} 

\affiliation{Department of Physics, University of Toronto, Toronto,
Ontario M5S 1A7, Canada} 

\date{\today} 

\begin{abstract}
The Josephson oscillations between condensates in a double-well trap
are known theoretically to be strongly effected by the mean field
interaction in dilute atomic gases. The most important effect is that
the amplitude of oscillation in the relative population of the two
wells is greatly suppressed due to the mean field interaction, which
can make it difficult to observe the Josephson effect. Starting from
the work of Raghavan, Smerzi, Fantoni, and Shenoy, we calculate the
maximum amplitude of oscillation in the relative population as a
function of various physical parameters, such as the trap aspect
ratio, the Gaussian barrier height and width, and the total number of
atoms in the condensate. We also compare results for ${}^{23}$Na and
${}^{87}$Rb. Our main new result is that the maximum amplitude of
oscillation depends strongly on the aspect ratio of the harmonic trap
and can be maximized in a ``pancake'' trap, as used in the experiment
of Anderson and Kasevich. 
\end{abstract}

\pacs{03.75.Fi.~~05.30.Jp~~32.80.Pj}

\maketitle 

\section{Introduction}
In this paper we consider a Bose-Einstein condensate (BEC) at zero
temperature trapped in a double-well potential, which is created by
superimposing a harmonic trap and a Gaussian barrier. With a finite
barrier height that is greater than the condensate chemical potential,
the Gross-Pitaevskii dynamics of the condensate can be well described
as a superposition of right and left localized condensates, the
so-called two-mode ansatz~\cite{Raghavan1999b}. The weak link allows
Josephson tunneling of condensate atoms between wells when there is a
chemical potential difference (or a difference in phases) between the
wells.

A double well system was implemented in the experiment described in
Ref.~\cite{Andrews1997b}, where the Gaussian barrier was created by
focusing a blue-detuned far-off-resonant light sheet into the center
of a magnetic trap. The trap had a ``cigar'' geometry, that is
$\lambda \gg 1$, where we have defined the aspect ratio of the radial
and axial trap frequencies $\lambda \equiv \omega_\rho/\omega_z$,
which turns out to play an important role in our calculations. In the
experiment of Ref.~\cite{Andrews1997b}, an {\emph{infinite}} barrier
along the $z$-axis was used to separate two independent condensates of
about five million ${}^{23}$Na atoms and then removed to allow the
condensates to interfere. In this paper we focus on Josephson
tunneling through a {\emph{finite}} barrier, which was not reported in
Ref.~\cite{Andrews1997b}. To date, there have been no reports of
experimental observations of Josephson tunneling in a double-well
trap. Josephson tunneling of a condensate in a one dimensional optical
lattice was reported in Ref.~\cite{Anderson1998a}. Each well in the
array held a condensate of about one thousand ${}^{87}$Rb atoms. The
chemical potential offset due to gravity caused the condensates to
tunnel out of the wells. A crucial characteristic of the experiment is
that the wells had an extremely ``pancake'' geometry ($\lambda \ll 1$)
with $\lambda \approx 0.002$~\cite{Anderson1998a,AndersonPrivate}, in
contrast to the ``cigar'' geometry of the trap used in
Ref.~\cite{Andrews1997b}.  A related system of two hyperfine states
coupled by applied electromagnetic fields was reported in
Refs.~\cite{Hall1998b,Matthews1999a,Matthews1999b} for the case of
strong coupling. Tunneling in a three-component condensate was
reported in Ref.~\cite{Miesner1999a}.

The double-well tunneling system for a Bose condensate has been
treated extensively in the recent theoretical
literature~\cite{Javanainen1986a,Milburn1997a,Zapata1998a,Raghavan1999b,Salasnich1999a}. Nonlinear
Josephson equations describing the oscillations of the relative
population $\eta(t)= (N_1(t)-N_2(t))/N_c$ and phase $\phi(t) =
\phi_1(t)-\phi_2(t)$ between the wells have been thoroughly
explored~\cite{Raghavan1999b,Marino1999a}, and this analysis will be
the basis of our paper. Here, the indices $1$ and $2$ label the left
and right wells and $N_c = N_1 + N_2$ is the fixed total condensate
population, so that $-1 \leq \eta(t) \leq 1$. A crucial property of
this system is that, due to the nonlinear terms in the Josephson
equations arising from the mean field interaction, there is a maximum
amplitude of oscillation $\eta_{\rm{max}} < 1$ less than unity. If the
initial population difference $\eta(0)\gtrsim\eta_{\rm{max}}$, then
very little tunneling occurs, a phenomenon referred to as macroscopic
quantum ``self-trapping''~\cite{Raghavan1999b}. The calculations of
Ref.~\cite{Salasnich1999a} were carried out using realistic parameters
corresponding to the experiment of Ref.~\cite{Andrews1997b}, but used
a finite barrier height to allow tunneling to occur. An extremely
small value of $\eta_{\rm{max}} \ll 0.01$ was found, indicating that
Josephson oscillations in that system would be very difficult to
observe due to experimental error in measuring $\eta(t)$ (which we
assume to be of the order of a few percent).

Because of the importance of having clear experimental demonstrations
of quantum phase oscillations, we feel it is useful to explore a wider
range of physical parameters in order to find optimal conditions for
observing Josephson oscillations in a double-well. This is the main
purpose of the present paper. The key quantities that determine
$\eta_{\rm{max}}$ are the condensate density, the s-wave scattering
length $a$, the atomic mass $m$, and the coupling strength between
wells. In the absence of interactions ($a=0$), the problem reduces to
a simple linear two-level system, so that $\eta_{\rm{max}}=1$. The
effect of the mean field interaction can be reduced by reducing the
density, scattering length, or the mass. In this paper, we do not vary
$m$ or $a$, but rather, we consider two types of atoms, ${}^{23}$Na
and ${}^{87}$Rb, each with a fixed mass and scattering length. We
calculate $\eta_{\rm{max}}$ for a wide range of trap parameters by
varying the trap geometry determined by $\omega_z$ and $\omega_\rho$,
the barrier geometry, and the number of condensate atoms $N_c$.  We
find $\eta_{\rm{max}}$ is optimized by making the trap more
``pancake'' shaped ($\lambda \ll 1$), which has the effect of
maximizing the weak link area, while maintaining a sufficiently low
condensate density. This result is consistent with the fact that
Josephson tunneling was clearly observed in Ref.~\cite{Anderson1998a}
for an array of wells possessing an extreme ``pancake'' geometry. This
behavior is in contrast to that for a trap with a ``cigar'' geometry,
which minimizes the weak link area and compresses the condensate in
two dimensions rather than one, resulting in a greater density.

\section{Derivation of Model}

At zero temperature, the condensate dynamics is well described by the
Gross-Pitaevskii (GP) equation \beq i\hbar {\partial \Phi(\br,t) \over
\partial t} = \big [ H_0 + g n_c(\br,t) \big] \Phi(\br,t) ,
\label{TDGP} 
\eeq where $n_c(\br,t)=|\Phi(\br,t)|^2$ is the condensate density and
\beq H_0 =-{\hbar^2 \over 2 m} \nabla^2 + U_{\rm{ext}}(\br)
. \label{H0}\eeq Here, the external potential
$U_{\rm{ext}}(\br)=U_H(\br)+U_B(\br)$ is created by superimposing a
harmonic trap \beq U_H(\rho,z) = m \omega_z^2(\lambda^2 \rho^2 +
z^2)/2, \eeq and a Gaussian barrier along the $z$ axis. \beq U_B(z) =
U_0 \exp({-z^2/2\sigma^2}).\label{UB}\eeq In our calculations, we
consider a range of values for the trap frequencies $\omega_z$ and
$\omega_\rho$, the barrier height $U_0$ and width $\sigma$, and
the condensate population $N_c$.

\subsection{Two-mode ansatz}

For a barrier height that is sufficiently larger than the condensate
chemical potential \beq U_0 > \mu_c, \eeq there exist two nearly
degenerate stationary solutions of the GP equation \beq \big [ H_0 + g
n_c(\br) \big] \Phi_\alpha(\br) = \mu_\alpha \Phi_\alpha(\br),
\label{SGP} \eeq where $n_c(\br)=N_c|\Phi_\alpha(\br)|^2$. The index
$\alpha=\{S,A\}$ indicates the reflection symmetry along the $z$-axis,
symmetric or antisymmetric, respectively. These solutions satisfy the
orthonormality condition $\int
d\br\Phi_\alpha(\br)\Phi_\beta(\br)=\delta_{\alpha\beta}$. Left and
right localized states can be formed from these eigenstates
$\Phi_1(\br)=\big [ \Phi_S(\br)+ \Phi_A(\br) \big ]/\sqrt{2}$ and
$\Phi_2(\br)=\big [ \Phi_S(\br)- \Phi_A(\br) \big ]/\sqrt{2}$,
respectively, which are also orthonormal.

In the two-mode approximation, the condensate dynamics is described by
the following ansatz solution~\cite{Raghavan1999b} \beq \Phi(\br,t) =
\psi_1(t)\Phi_1(\br) + \psi_2(t)\Phi_2(\br). \label{2mode}\eeq Here,
the complex coefficients $\psi_i(t)=\sqrt{N_i(t)}\exp[i\phi_i(t)]$ are
spatially uniform and contain all of the time dependence, while the
two states $\Phi_1(\br)$ and $\Phi_2(\br)$ are localized in the left
and right wells, respectively, and contain all of the position
dependence.  In a more accurate theory, one could account for the slow
time evolution of the states $\Phi_i(\br)$ due to the mean field
interaction since the population $N_i(t)$ in each well evolves in
time. If the amplitude of oscillation is relatively small, this can be
shown to have a negligible effect~\cite{Salasnich1999a,Williams1999a}.

\subsection{Gaussian variational solution}
Since we are exploring the parameter space of $\{ \omega_z,
\omega_\rho, U_0, \sigma, N_c\}$ and must solve \eq{SGP} many times, it
is useful to construct a simplified model solution of the left and
right states $\Phi_i(\br)$.  Several earlier studies have shown that a
Gaussian variational solution gives an accurate description of the
condensate~\cite{Baym1996a,Perez-Garcia1996a,Menotti2001a}. We
therefore employ this procedure by taking the following
ansatz for $\Phi_\alpha(\br)$
\beq
\Phi_\alpha(\br) = C_\alpha e^{-A_\alpha^2 \rho^2}\big [
e^{-B_\alpha^2 (z+z_\alpha)^2} \pm e^{-B_\alpha^2 (z-z_\alpha)^2}
\big ] . \label{gauss}\eeq
Here, $C_\alpha$ is a normalization constant given by
\beq
C_\alpha = \sqrt {2^{1/2} A_\alpha^2 B_\alpha \over
{\pi^{3/2} \big [ 1 \pm \exp(-2 B_\alpha^2 z_\alpha^2) \big ]}}.
\label{C}\eeq
The three variational coefficients $\{A_\alpha, B_\alpha, z_\alpha \}$
must be determined by minimizing the Gross-Pitaevskii energy
functional \beq E(\Phi) = \int d \br \Big[ {\hbar^2 \over {2 m}}
|\nabla \Phi|^2 + U_{\rm{ext}} |\Phi|^2 + {g\over 2} |\Phi|^4 \Big ]
\label{E}
\eeq with respect to variations in $\{A_\alpha, B_\alpha, z_\alpha
\}$. We note that these coefficients will be different for the
symmetric and anti-symmetric states. The Gaussian form of our
variational ansatz permits us to evaluate the integral in \eq{E}
analytically. However, due to the nonlinear terms in the resulting
expression, we carry out the minimization of $E(\Phi)$ numerically. In
Section III, we compare the Gaussian variational solution with the
full numerical solution of \eq{SGP} and find qualitative agreement
over a wide range of parameters.

\subsection{Equations for the relative phase and population}

We substitute the two-mode ansatz \eq{2mode} into the GP equation
\eq{TDGP}, multiply by $\Phi_i(\br)$ and integrate over position to
obtain equations of motion for the complex coefficients
$\psi_i(t)$~\cite{Raghavan1999b} \bea i \hbar \dot\psi_i(t) &=& \big
[E_i + U_i N_i(t)\big ] \psi_i(t) \nonumber \\ &-& \big[K + U_{1,12}
N_1(t) + U_{2,12} N_2(t) \big ] \psi_j(t),
\label{psidot}
\eea where we have used the orthonormality property.  Here, $E_i$ is
the bare zero-point energy in well $i$, \beq E_i = \int d\br
\Phi_i(\br) H_0 \Phi_i(\br) \label{Ei}\eeq and $U_i N_i(t)$ is the
mean-field interaction energy, with \beq U_i = g \int d\br
\Phi_i^4(\br).  \label{Ui}\eeq The term $K$ in \eq{psidot} is the
coupling energy between condensates due to the finite probability of
tunneling through the barrier, \beq K = -\int d\br \Phi_1(\br) H_0
\Phi_2(\br).
\label{K} \eeq In \eq{psidot}, we also include the mean-field
contribution to the coupling, given by $[U_{1,12} N_1(t) + U_{2,12}
N_2(t)]$, where \beq U_{i,12} = -g \int d\br
\Phi_i^2(\br)\Phi_1(\br)\Phi_2(\br). \label{Ui12} \eeq As shown
recently in Ref.~\cite{Zhang2001a}, these factors \eq{Ui12} give rise
to a time-dependent coupling term, but were neglected in the
calculation of Ref.~\cite{Raghavan1999b}. The integrals in
\eq{Ei}-\eq{Ui12} can be carried out analytically using the Gaussian
variational solution \eq{gauss}.

It is straightforward to obtain the equations of motion for the
phases and populations of each condensate by substituting $\psi_i(t) =
\sqrt{N_i(t)}e^{i \phi_i(t)}$ into \eq{psidot}, \bea \label{Ndoti}
\dot N_i(t) &=& {1
\over \hbar} \Big[U_{1,12} N_1(t) + U_{2,12} N_2(t) \nonumber \\ &+& K
\Big ] [N_i(t) N_jt)]^{1/2} \sin[\phi_i(t) - \phi_j(t)] \\ 
\hbar \dot\phi_i(t)  &=& -E_i - U_i N_i(t) + \Big[U_{1,12}
N_1(t) + U_{2,12} N_2(t) \nonumber \\ &+& K \Big ] \Big ({N_j(t) \over
N_i(t)} \Big )^{1/2} \cos[\phi_i(t) - \phi_j(t)] .
\label{phidoti}
\eea 
Using \eq{Ndoti} and \eq{phidoti}, we can derive two equations for the
relative phase $\phi(t) = \phi_1(t) - \phi_2(t)$ and the relative
population $\eta(t) = [N_1(t) - N_2(t)]/N_c$~\cite{Raghavan1999b} \bea
\label{etadot}
\dot\eta(t) &=&\omega_J(t) \sqrt{1-\eta^2(t)} \sin{\phi(t)}, \nonumber
\\ \\
\label{phidot} \dot\phi(t) &=& -\Delta \omega -
\omega_C \eta(t) - \omega_J(t)
{\eta(t)\cos{\phi(t)}\over\sqrt{1-\eta^2(t)}}, \eea Here we have
simplified the notation by defining the Josephson frequency \beq
\omega_J(t) \equiv \big [ 2K + N_c U_{1,12}(1+\eta(t)) + N_c
U_{2,12}(1-\eta(t))\big ]/\hbar, \eeq the ``capacitive'' frequency
proportional to the mean field interaction \beq \omega_C \equiv
N_c(U_1 + U_2)/2\hbar, \eeq and the frequency due to the difference in
zero-point energies \beq \Delta \omega \equiv \big [E_1 - E_2 +
(U_1-U_2)N_c/2 \big ]/\hbar .\eeq

For simplicity, we restrict our calculations to a symmetric
double-well, since the overall behavior of $\eta_{\rm{max}}$ does not
depend much on a relative offset in the well depths. In this case, the
spatially averaged quantities \eq{Ei}, \eq{Ui}, and \eq{Ui12} are
equal for each well ($E_1=E_2$, $U_1 = U_2$,
$U_{1,12}=U_{2,12}\equiv U_{12}$), allowing us to make the
simplifications $\Delta \omega = 0$ and \beq \omega_J = 2(K + N_c
U_{12})/\hbar , \label{omegaJ2}\eeq which is now
time-independent. Using these results, the equations of motion
\eq{etadot} and \eq{phidot} reduce to \bea \dot \eta(t) &=& \omega_J
\sqrt{1 - \eta^2(t)} \sin \phi(t) ,\label{etadot2} \\ \dot \phi(t)&=&
-\omega_C \eta(t) - \omega_J {\eta(t) \over \sqrt{1 - \eta^2(t)}} \cos
\phi(t) .\label{phidot2} \eea

In Ref.~\cite{Raghavan1999b}, Raghavan {\emph{et al.}} give an
analytic solution of \eq{etadot2} and \eq{phidot2} in terms of
Jacobian elliptic functions. The solution depends on the initial
conditions for the relative number and phase, $\eta(0)$ and
$\phi(0)$. From the solution for $\eta(t)$ we can define two maximum
amplitudes of oscillation, one for the case $\eta(0)=0$, the other in
the case of a finite $\eta(0)\neq 0$.  For equal populations
initially, i.e. $\eta(0)=0$, the maximum amplitude of oscillation
$\eta_{\rm{max}}$ can be obtained from equation (B5a) of
Ref.~\cite{Raghavan1999b} by taking $\phi(0)=\pi/2$, which gives \beq
\eta_{\rm{max}} = {1\over \Lambda} \Big [2\big(\sqrt{\Lambda^2+1}-1
\big ) \Big ]^{1/2},
\label{etamax}\eeq where $\Lambda \equiv \omega_C/\omega_J$. In the 
non-interacting limit ($g=0$), $\omega_C=0$ and hence $\Lambda = 0$,
so that \eq{etamax} gives $\eta_{\rm{max}} = 1$, as it must in this
limit. In the strongly interacting limit $\Lambda \gg 1$, \eq{etamax}
takes the approximate form $\eta_{\rm{max}} \approx \sqrt{2(\Lambda -
1)}/\Lambda \ll 1$. If instead the initial population difference is
nonzero, then a slightly higher amplitude $\eta_c$ can be obtained
compared to that given in \eq{etamax}. Taking $\phi(0)=0$ in equation
(4.9) of Ref.~\cite{Raghavan1999b}, we obtain \beq \eta_c = 2
\sqrt{\Lambda - 1} / \Lambda .\label{etac} \eeq If $\eta(0) > \eta_c$,
then the solution is self-trapped. We note that in the strongly
interacting limit $\Lambda \gg 1$, these two quantities are
approximately equal and satisfy the following relation $\eta_c \simeq
\sqrt 2 \eta_{\rm{max}}$. We work with $\eta_{\rm{max}}$ in our
calculations rather than $\eta_c$ because \eq{etamax} is valid over
the whole range of values for $\Lambda$.

For small-amplitude oscillations, \eq{etadot2} and \eq{phidot2} can be
linearized and combined to give \beq \ddot\eta(t) + \omega_J (\omega_C
+ \omega_J)\eta(t) = 0. \label{etalinear}\eeq This equation tells us
that the system oscillates at the ``Josephson plasma'' frequency \beq
\omega_{\rm{JP}} = \sqrt{\omega_J(\omega_C+ \omega_J)}.  \eeq In the
case where the mean field interaction is large compared to the
coupling energy $\omega_C/\omega_J \gg 1$, this reduces to
$\omega_{\rm{JP}} =\sqrt{\omega_J\omega_C}$.  Due to the mean field
interaction, the actual frequency of oscillation between the two wells
can be large $\omega_{\rm{JP}} > \omega_z$, even though the coupling
is weak $\omega_J < \omega_z$.

\subsection{Illustration of nonlinear oscillations}
It is instructive to show the time evolution of $\eta(t)$ in order to
visualize the behavior of the nonlinear Josephson oscillations and to
clarify the precise meaning of $\eta_{\rm{max}}$, $\eta_c$, and
$\omega_{\rm{JP}}$. In Fig. 1 we graph $\eta(t)$ for three different
initial conditions $\eta(0)$ and $\phi(0)$ for the solution of the
coupled nonlinear equations \eq{etadot2} and \eq{phidot2}. We first
consider the case of equal populations $\eta(0)=0$ and an initial
relative phase of $\phi(0)=\pi/2$, shown by the solid line. The
population difference $\eta(t)$ oscillates with an amplitude of
$\eta_{\rm{max}}$. In the non-interacting limit
(eg. $\omega_C\rightarrow 0$), the problem reduces to a linear coupled
two-level system and a complete population transfer between wells
would occur in this case, $\eta_{\rm{max}}=1$, at a frequency of
$\omega_J$. The amplitude suppression introduced by the mean field
interaction arises from the term $-\omega_C \eta(t)$ in \eq{phidot2},
which acts as a time-dependent offset in the zero-point energies of
the two wells.  We also indicate the period of small-amplitude
oscillation $T_{\rm{JP}} \equiv 2\pi/\omega_{\rm{JP}}$ on the graph,
which is valid when the amplitude is much less than $\eta_{\rm{max}}$
(we have verified this in calculations not shown here). $T_{\rm{JP}}$
gives a useful estimate of the period, which, in general, depends on
the initial conditions.

We also consider the case of unequal initial populations $\eta(0)\neq
0$. For the dashed line in Fig. 1, we have taken $\eta(0)=0.9\eta_c$
and $\phi(0)=0$, where the critical amplitude is $\eta_c=0.11$ for
this case. If the initial population difference is larger than
$\eta_c$, then the population becomes ``trapped'', as shown for the
case of $\eta(0)=0.15$ by the dot-dashed line. It is important to
realize that in the non-interacting limit ($\omega_C \rightarrow 0$),
this self-trapping does not occur. In the absence of interactions, a
full oscillation of the dot-dashed line would occur between $\eta =\pm
0.15$, at a frequency of $\omega_J$. We remark that for $\eta(0)
\approx \eta_c$, higher harmonics appear in the solution due to the
nonlinearity~\cite{Raghavan1999b}. We refer the reader to
Ref.~\cite{Raghavan1999b} for a more detailed discussion of this
system, where a full range of solutions of \eq{etadot2} and
\eq{phidot2} is explored. We also note that this same behavior is also
found in Ref.~\cite{Williams1999a} for a condensate with two internal
states coupled by a weak external electromagnetic field.
\begin{figure}
  \centerline{\epsfig{file=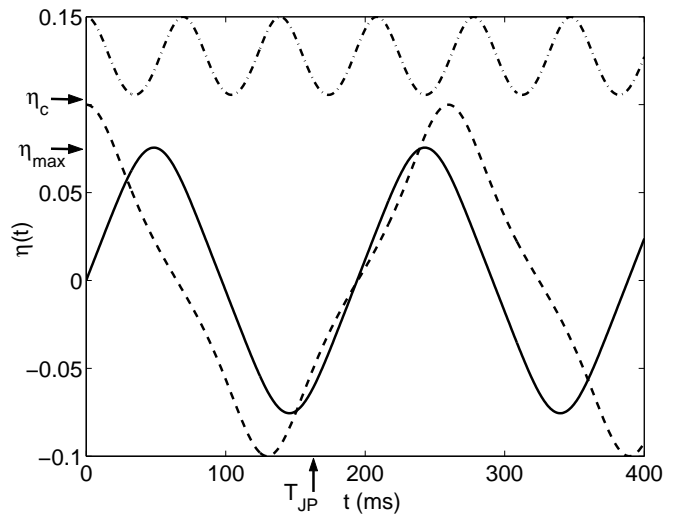,height=2.7in}}
\caption{Amplitude-suppressed Josephson oscillations of $\eta(t)$. The
nonlinear equations \eq{etadot2} and \eq{phidot2} for $\eta(t)$ and
$\phi(t)$ are solved for different initial conditions: $\eta(0)=0$,
$\phi(0)=\pi/2$ (solid), $\eta(0)=0.1$, $\phi(0)=0$ (dashed),
$\eta(0)=0.15$, $\phi(0)=0$ (dot-dashed). We indicate along the
vertical axis the critical value $\eta_c=0.11$ and the maximum
amplitude $\eta_{\rm{max}}=0.075$ obtained when $\eta(0)=0$. Along the
horizontal axis we also indicate the estimated period of the
Josesphson oscillations $T_{\rm{JP}}\equiv 2\pi/\omega_{\rm{JP}}$.
The physical parameters used correspond to the trap of
Ref.~\cite{Andrews1997b} in Table 1, but for $N_c = 1000$ atoms.}
\end{figure}

\section{Results}
In this section we calculate $\eta_{\rm{max}}$ as given by \eq{etamax}
as a function of the trap frequencies $\omega_z/2\pi$ and
$\omega_\rho/2\pi$, the barrier height $U_0$ and width $\sigma$, and
the condensate population $N_c$. To do this, we solve for the left and
right localized states $\Phi_i(\br)$ and then calculate the quantities
\eq{Ei}-\eq{Ui12}, so that the frequencies $\omega_J$ and $\omega_C$
can then be computed. We compare two different solutions of
$\Phi_i(\br)$, a direct numerical solution~\cite{Holland1997a} of the
stationary GP equation \eq{SGP} and the Gaussian variational ansatz
\eq{gauss}. The variational coefficients
$\{A_\alpha,B_\alpha,z_\alpha\}$ are computed numerically by
minimizing the energy functional $E(\Phi)$ \eq{E} using the
Nelder-Mead simplex search algorithm in the software package MATLAB.
\begin{table}
\caption{Table of trap parameters. In the calculation of
$\eta_{\rm{max}}$, we take $N_c = 10^4$, $\chi=1.5$, and $\sigma
= 1\mu$m. For the scattering lengths of ${}^{23}$Na and ${}^{87}$Rb,
we take $a_{\rm{Na}}=3$ nm and $a_{\rm{Rb}}=5.7$ nm.}
\vspace{0.3cm}
{\centering \begin{tabular}{|c|cccc|c|}
\hline 
trap&
\( \omega _{z}/2\pi  \) (Hz)&
\( \omega _{\rho }/2\pi  \) (Hz)&
\( \,\,\,\,\lambda\,\,\,\,  \)&
\,\,\,atom\,\,\,&
\( \,\,\,\,\eta_{\rm{max}}\,\,\,\,  \) \\
\hline 
\hline 
Ref.~\protect\cite{Andrews1997b}&
19&
250&
13&
${}^{23}$Na&
0.036 \\
\hline 
Ref.~\protect\cite{Matthews1999a}&
65&
24&
0.37&
${}^{87}$Rb&
0.14 \\
\hline 
Ref.~\protect\cite{Matthews1999b}&
8&
8&
1&
${}^{87}$Rb&
0.26 \\
\hline 
Ref.~\protect\cite{Anderson1998a}&
7200&
17&
0.002&
${}^{87}$Rb&
0.85 \\
\hline 
\end{tabular}\par}
\vspace{0.3cm}
\end{table}

While varying the physical parameters $\omega_z$, $\omega_\rho$,
$\sigma$, and $N_c$ in our calculations, we keep $U_0$ at a fixed
height relative to the chemical potential $\mu_S$, which also
varies. The bottom of the double well $U_{\rm{min}}$ is obtained by
minimizing $U_{\rm{ext}}$ along the $z$-axis to give\beq U_{\rm{min}}
= m \omega_z^2 \sigma^2 \big [1 - \ln(m \omega_z^2 \sigma^2/U_0) \big
].  \eeq In our calculations, we define the ratio $\chi$ of the
barrier height $U_0$ to the chemical potential $\mu_S$ of the
symmetric solution, each measured relative to the bottom of the well
\beq \chi = {(U_0 - U_{\rm{min}}) \over {(\mu_S -
U_{\rm{min}})}}.\label{chi} \eeq We emphasize that in {\emph{all}} of
our calculations, the coupling due to tunneling is {\emph{weak}}, with
$0.05 \lesssim \omega_{\rm{JP}}/\omega_z \lesssim 0.5$.

We first calculate $\eta_{\rm{max}}$ for a few typical magnetic traps
shown in Table 1. For the ``cigar'' trap of Ref.~\cite{Andrews1997b},
taking $N_c = 5 \times 10^6$ and a barrier width of $\sigma = 6\mu$m,
the maximum amplitude of oscillation is miniscule $\eta_{\rm{max}}=5.2
\times 10^{-6}$. Even if we greatly reduce the condensate density by
lowering the population $N_c = 10^4$ and increase the coupling by
reducing the barrier width $\sigma = 1 \mu$m, the maximum amplitude is
still very small, with $\eta_{\rm{max}}=0.036$. The Josephson
oscillations would not be visible in this case if the experimental
error in measuring $\eta_{\rm{max}}$ were a few percent. For the
``pancake'' trap of Ref.~\cite{Matthews1999a}, taking a modest
condensate population $N_c = 10^4$ and a narrow barrier $\sigma = 2
\mu$m, we find $\eta_{\rm{max}} = 0.054$. A better result can be
obtained by decreasing the barrier width $\sigma = 1 \mu$m, which
gives $\eta_{\rm{max}} = 0.14$. Finally, we consider
the spherical trap of Ref.~\cite{Matthews1999b}, which is a much
shallower trap than the other two, so that the density of the
condensate is much less. Taking $N_c = 10^4$ and $\sigma = 2\mu$m, we
find $\eta_{\rm{max}} = 0.12$, and for $\sigma = 1\mu$m,
$\eta_{\rm{max}} = 0.26$. For the optimistic parameters of this
shallow trap, Josephson oscillations should be observable.

For comparison we also consider the experimental setup of
Ref.~\cite{Anderson1998a}, where a condensate was loaded into a one
dimensional optical lattice. Each well in the array can be modeled as
being harmonic as a first approximation, for which one obtains
$\omega_z/2\pi \sim 7200$ Hz and $\omega_\rho/2\pi \sim 17$
Hz~\cite{Anderson1998a,AndersonPrivate}. For a rough comparison, we
consider a double well with this geometry. Taking $\chi=1.5$, $\sigma
= 1 \mu$m, and $N_c = 10^4$ ${}^{87}$Rb atoms, we find
$\eta_{\rm{max}}=0.85$. This value increases to $\eta_{\rm{max}} =
0.99$ if we reduce the population to $N_c=10^3$.

\begin{figure}
  \centerline{\epsfig{file=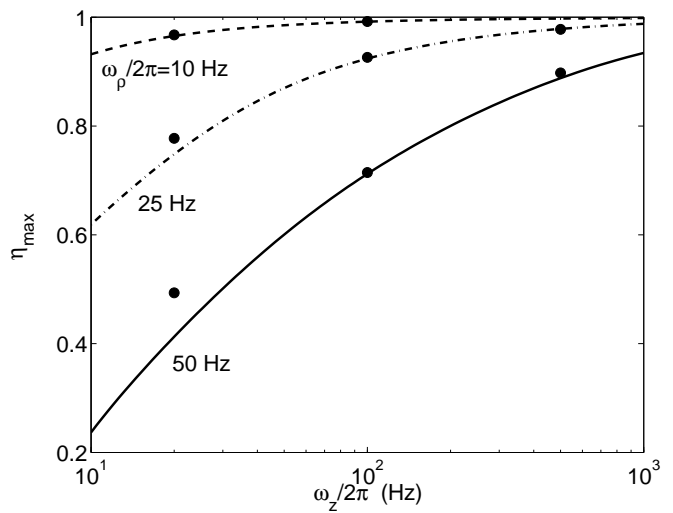,height=2.7in}}
\caption{$\eta_{\rm{max}}$ as a function of $\omega_z/2\pi$ for three
values of $\omega_\rho/2\pi$. The three lines are obtained using the
Gaussian variational solution \eq{gauss} and the filled circles are
obtained from a full numerical solution of the GP equation
\eq{SGP}. In this calculation, the condensate population is held fixed
at $N_c = 10^3$ atoms of ${}^{23}$Na. The relative barrier
height~\eq{chi} and width as given in~\eq{UB} are fixed at $\chi =
1.5$ and $\sigma=1 \mu$m, respectively. }
\end{figure}
Motivated by these results, we obtain a very crude estimate of the
scaling of $\eta_{\rm{max}}$ with the parameters. In the strongly
interacting limit $\Lambda \gg 1$, \eq{etamax} reduces to \beq
\eta_{\rm{max}} \sim {1\over\sqrt{\Lambda}} \sim \sqrt{\omega_J \over
\omega_C}. \label{etamax_approx} \eeq Taking a simple WKB approach to
calculate the tunneling probability~\cite{Zapata1998a,Sols1999a}, an
approximate scaling of the Josephson frequency can be obtained
$\omega_J/\omega_z \sim \exp[-\sqrt{m \sigma^2
(U_0-\mu)/\hbar^2}]$. One can also get the approximate scaling of
$\omega_C$ on $m$ and $a$ by taking $\omega_C \sim \mu_{\rm{TF}}$,
where $\mu_{\rm{TF}}/\hbar\omega_z = 0.5[15 \lambda^2 N_c
a/z_{\rm{ho}}]^{2/5}$ is the chemical potential of a condensate in a
harmonic trap in the Thomas-Fermi limit~\cite{Dalfovo1999a}. Inserting
the definition of the oscillator length, $z_{\rm{ho}} = \sqrt{\hbar/m
\omega_z}$, the expression can be simplified to $\omega_C/\omega_z
\sim (m a^2 N_c^2\lambda^4 \omega_z/\hbar)^{1/5}$. Inserting these
crude estimates of $\omega_J$ and $\omega_C$ into \eq{etamax_approx},
we find that $\eta_{\rm{max}}$ scales like \beq \eta_{\rm{max}} \sim
{\exp[-\sqrt{m \sigma^2 (U_0-\mu)/4\hbar^2}]\over{(m a^2
\lambda^4 N_c^2 \omega_z/\hbar)^{1/10}}}. \label{etamax_scale} \eeq This indicates that the amplitude
of oscillation can be increased by decreasing $m$, $a$, $N_c$,
$\sigma$, or $U_0$. For a fixed $\omega_z$, $\eta_{\rm{max}}$
increases as $\lambda$ is decreased, while for a fixed aspect ratio
$\lambda$, $\eta_{\rm{max}}$ decreases when the trap is tightened by
increasing $\omega_z$.  Expression $\eq{etamax_scale}$ only gives a
very crude estimate of the scaling behavior. In the following
discussion, we explore the dependence of $\eta_{\rm{max}}$ on the
physical parameters more quantitatively by calculating $\omega_J$ and
$\omega_C$ numerically.

\begin{figure}
  \centerline{\epsfig{file=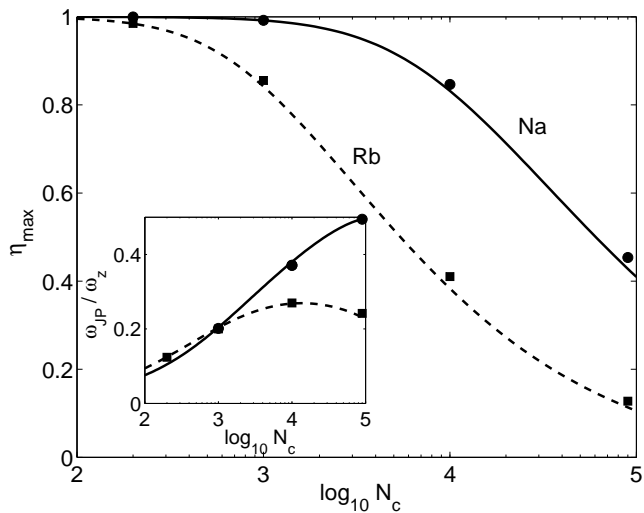,height=2.7in}}
\caption{$\eta_{\rm{max}}$ as a function of the condensate population
$N_c$ for ${}^{23}$Na and ${}^{87}$Rb. The lines are obtained using
the Gaussian variational solution \eq{gauss} and the filled circles
(${}^{23}$Na) and squares (${}^{87}$Rb) are obtained from a full
numerical solution of the GP equation \eq{SGP}. In the inset we show
how the frequency of oscillation $\omega_{\rm{JP}}/\omega_z$ varies
with $N_c$. In this calculation, the harmonic trap frequencies are
held fixed at $\nu_z=100$ Hz and $\nu_\rho=10$ Hz. The relative
barrier height and width are fixed at $\chi = 1.5$ and $\sigma=1
\mu$m, respectively.}
\end{figure}
In Figs. 2 - 5 we plot $\eta_{\rm{max}}$ (given by \eq{etamax}) as a
function of the trap geometry (by varying $\omega_z$ and
$\omega_\rho$), $N_c$, and the barrier height $\chi$ and width
$\sigma$. In each figure, we show results for the Gaussian variational
solution (lines) and the direct numerical solution of the stationary
GP equation \eq{SGP} (filled points). The Gaussian variational
solution agrees with the full solution within a few percent over a
broad range of parameters. The greatest discrepancy occurs in Fig. 2
as $\omega_z$ decreases at fixed $\omega_\rho$ and in Fig. 4 as
$\sigma$ increases, with the two solutions differing by a factor of 2
at $\sigma=4\mu$m for ${}^{87}$Rb. In all cases, the Gaussian solution
slightly underestimates $\eta_{\rm{max}}$.

Our most important result is shown in Fig. 2, which shows the strong
dependence of $\eta_{\rm{max}}$ on the trap geometry. For a fixed
value of $\omega_\rho$, $\eta_{\rm{max}}$ can be optimized by
increasing the axial trap frequency $\omega_z$, that is, by making
the trap more ``pancake'' like. In contrast, for a fixed value of
$\omega_z$, $\eta_{\rm{max}}$ decreases as $\omega_\rho$ is
increased. It is important to realize that both quantities $\omega_C$
and $\omega_J$ are effected by changing the aspect ratio of the trap.
By making the trap more ``pancake'' like, one is increasing the area
of contact between the two wells, which increases $\omega_J$. On the
other hand, compressing the condensate in the $z$ dimension also leads
to a higher density, causing $\omega_C$ to increase as well. However,
the ratio of these two quantities in \eq{etamax_approx} increases for
increasing $\omega_z$. In contrast, by making the trap more
``cigar'' like, one is decreasing the area of contact between wells,
resulting in a decrease in $\omega_J$ as $\omega_\rho$
increases. Furthermore, in this case one is compressing the condensate
in two dimensions rather than one, which results in a more dramatic
increase in the density and a corresponding increase in $\omega_C$, so
that the ratio $\omega_J/\omega_C$ decreases with increasing
$\omega_\rho$.

\begin{figure}
  \centerline{\epsfig{file=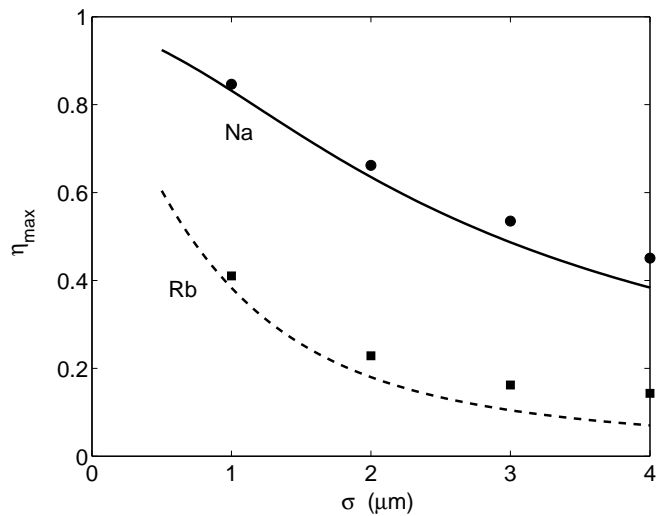,height=2.7in}}
\caption{$\eta_{\rm{max}}$ as a function of the barrier width $\sigma$
for ${}^{23}$Na and ${}^{87}$Rb. In this calculation, the harmonic
trap frequencies are held fixed at $\nu_z=100$ Hz and $\nu_\rho=10$
Hz. The relative barrier height is fixed at $\chi = 1.5$ and the
population is $N_c = 10^4$ atoms.}
\end{figure}
In Fig. 3 we vary the condensate population $N_c$ from $10^2$ to
$10^5$ atoms, and show $\eta_{\rm{max}}$ for both ${}^{23}$Na and
${}^{87}$Rb. We see in Figs. 2 - 4 that a much larger amplitude of
oscillation $\eta_{\rm{max}}$ can be achieved using ${}^{23}$Na
compared to ${}^{87}$Rb. This behavior is expected based on the
rough scaling of $\eta_{\rm{max}}$ given in \eq{etamax_scale}. We can
understand why ${}^{23}$Na is favorable over ${}^{87}$Rb, since the
mass $m$ of ${}^{23}$Na is a about a factor of four smaller than that
of ${}^{87}$Rb, and the scattering length $a$ of ${}^{23}$Na is almost
half that of ${}^{87}$Rb. As $N_c$ increases in Fig. 3, the effect of
the mean field interaction becomes more pronounced causing $\omega_C$
to increase, while this has only a small effect on $\omega_J$. We
also show the dependence of $\omega_{\rm{JP}}$ on the condensate
number $N_c$ in the inset.

We also show the dependence of $\eta_{\rm{max}}$ on the Gaussian
barrier parameters by varying the relative height $\chi$ \eq{chi} in
Fig. 4 and the width $\sigma$ in Fig. 5. We note that varying
these two parameters has very little effect on $\omega_C$, which
remains roughly constant in these plots. The behavior of
$\eta_{\rm{max}}$ is not surprising: as the barrier height and width
are each increased, the coupling between wells is diminished so that
$\omega_J$ decreases, resulting in a decrease in $\eta_{\rm{max}}$
as indicated by \eq{etamax_approx}.
\begin{figure}
  \centerline{\epsfig{file=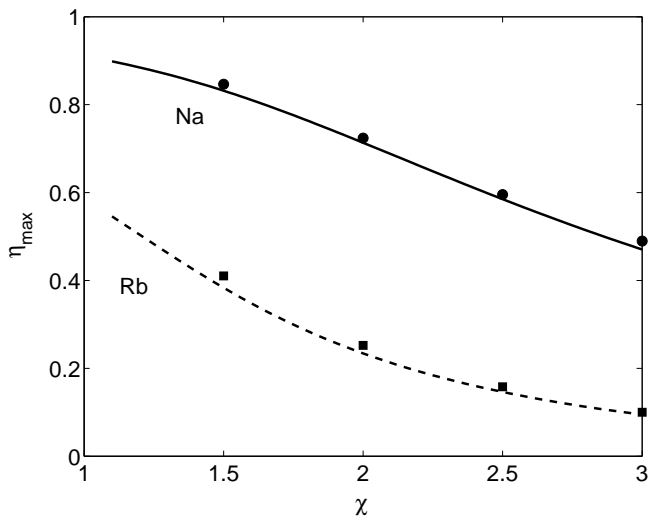,height=2.7in}}
\caption{$\eta_{\rm{max}}$ as a function of the relative barrier
height $\chi$ defined in~\eq{chi} for ${}^{23}$Na and ${}^{87}$Rb. In
this calculation, the harmonic trap frequencies are held fixed at
$\nu_z=100$ Hz and $\nu_\rho=10$ Hz. The barrier width is fixed at
$\sigma=1 \mu$m and the population is $N_c = 10^4$ atoms.}
\end{figure}

\section{Conclusions}
In this paper we have focussed on an important property of the
double-well Josephson system for a dilute atomic Bose-Einstein
condensate: the maximum amplitude of oscillation $\eta_{\rm{max}}$ of
the population between wells is suppressed due to the mean field
interaction~\cite{Milburn1997a,Zapata1998a,Raghavan1999b,Salasnich1999a}.
For typical magnetic trap parameters, we have shown that
$\eta_{\rm{max}}$ is very small, indicating that Josephson
oscillations may be visible only for the very shallow trap used in the
experiments described in~\cite{Matthews1999b}. We have shown that much
larger values of $\eta_{\rm{max}}$ can be achieved by decreasing the
aspect ratio $\lambda = \omega_\rho/\omega_z$, that is, by making the
trap more ``pancake'' like, in particular, for small values of the
radial frequency $\omega_\rho$. Our study was motivated by the fact
that Josephson oscillations were clearly observed in the experiment
reported in Ref.~\cite{Anderson1998a}, where a relatively small
condensate was loaded into an array of wells with an extremely small
aspect ratio $\lambda \approx
0.002$~\cite{Anderson1998a,AndersonPrivate}. We have also shown that
larger values of $\eta_{\rm{max}}$ can be attained by decreasing the
condensate population $N_c$, and decreasing the barrier height $\chi$
and width $\sigma$. Furthermore, conditions are much more favorable
for ${}^{23}$Na compared to ${}^{87}$Rb. It should be noted, however,
that present traps for ${}^{23}$Na atoms possess the less optimal
``cigar'' geometry.

We have restricted our calculations to zero temperature. Our study of
the optimal conditions for observing Josephson oscillations should be
extended to finite temperatures, where dissipative effects associated
with the thermal cloud of non-condensate atoms must be
included~\cite{Zapata1998a,Ruostekoski1998a}. Very few explicit
calculations of damping rates of Josephson oscillations in a dilute
Bose gas have been carried out, and further study of damping in
this system is needed.

I would like to thank Prof. A. Griffin for suggesting this problem and
for his useful comments. I thank B. Anderson for many insightful
comments and E. Zaremba, B. Jackson, and R. Spekkens for useful
discussions. This work was supported by a grant from NSERC.

\bibliography{bec}

\end{document}